\documentclass[conference]{IEEEtran}

\usepackage{array}
\usepackage{hyperref}
\usepackage{url}
\usepackage{graphicx}
\usepackage{epstopdf}
\usepackage{subcaption}
\usepackage{amsmath,amssymb,amsfonts,amsthm}
\usepackage{algorithm}
\usepackage{algorithmicx}
\usepackage{color}
\begin{document}

\title{FEAST: An Automated Feature Selection Framework for Compilation Tasks}

\author{
\IEEEauthorblockN{Pai-Shun Ting}
\IEEEauthorblockA{Department of Electrical Engineering and Computer Science\\
University of Michigan\\
Ann Arbor, Michigan 48109, USA\\
Email: paishun@umich.edu}
\\
\IEEEauthorblockN{Chun-Chen Tu}
\IEEEauthorblockA{Department of Statistics\\
University of Michigan\\
Ann Arbor, Michigan 48109, USA\\
Email: timtu@umich.edu}
\\
\IEEEauthorblockN{Pin-Yu Chen}
\IEEEauthorblockA{IBM T. J. Watson Research Center\\
Yorktown Heights, New York 10598, USA\\
Email: pin-yu.chen@ibm.com}
\\
\IEEEauthorblockN{Ya-Yun Lo}
\IEEEauthorblockA{Adobe Systems\\
	San Francisco, California 94103, USA\\
	Email: ylo@adobe.com}
\\
\IEEEauthorblockN{Shin-Ming Cheng}
\IEEEauthorblockA{Department of Computer Science and Information Engineering\\
	National Taiwan University of Sciecne and Technology\\
	Taipei 106, Taiwan\\
	Email: smcheng@mail.ntust.edu.tw}
}

\maketitle

\begin{abstract}
Modern machine-learning techniques greatly reduce the efforts required to conduct high-quality program compilation, which, without the aid of machine learning, would otherwise heavily rely on human manipulation as well as expert intervention. The success of the application of machine-learning techniques to compilation tasks can be largely attributed to the recent development and advancement of program characterization, a process that numerically or structurally quantifies a target program. While great achievements have been made in identifying key features to characterize programs, choosing a correct set of features for a specific compiler task remains an ad hoc procedure. In order to guarantee a comprehensive coverage of features, compiler engineers usually need to select excessive number of features. This, unfortunately, would potentially lead to a selection of multiple similar features, which in turn could create a new problem of bias that emphasizes certain aspects of a program's characteristics, hence reducing the accuracy and performance of the target compiler task. In this paper, we propose \textit{FEAture Selection for compilation Tasks} (\textit{FEAST}), an efficient and automated framework for determining the most relevant and representative features from a feature pool. Specifically, FEAST utilizes widely used statistics and machine-learning tools, including LASSO, sequential forward and backward selection, for automatic feature selection, and can in general be applied to any numerical feature set. This paper further proposes an automated approach to compiler parameter assignment for assessing the performance of FEAST. Intensive experimental results demonstrate that, under the compiler parameter assignment task, FEAST can achieve comparable results with about 18\% of features that are automatically selected from the entire feature pool. We also inspect these selected features and discuss their roles in program execution.

\end{abstract}


\IEEEpeerreviewmaketitle

\section{Introduction}

\linespread{0.7}
\begin{table}[!t]
	\linespread{1}
	\caption{List of all original 56 static features from cTuning Compiler Collection \cite{staticfeatures}.}
	\label{table_56_features}
	\centering
	\begin{tabular}{|m{0.3cm}|m{3.1cm}|m{0.3cm}|m{3.1cm}|}
		\hline
		ft1	& \# basic blocks in the method& ft29& \#  basic blocks with phi nodes in the interval [0, 3]\\ \hline
		ft2	& \# basic blocks with a single successor & ft30& \#  basic blocks with more then 3 phi nodes\\ \hline
		ft3	& \# basic blocks with two successors & ft31& \#  basic block where total \#  arguments for all phi-nodes is in greater then 5\\ \hline
		ft4	& \# basic blocks with more then two successors& ft32& \#  basic block where total \#  arguments for all phi-nodes is in the interval [1, 5]\\ \hline
		ft5 & \# basic blocks with a single predecessor&ft33& \#  switch instructions in the method\\ \hline
		ft6	& \# basic blocks with two predecessors& ft34& \#  unary operations in the method\\ \hline
		ft7	& \# basic blocks with more then two predecessors&ft35& \#  instruction that do pointer arithmetic in the method\\ \hline
		ft8	& \# basic blocks with a  single predecessor and a single successor&ft36& \#  indirect references via pointers ("*" in C)\\ \hline
		ft9	& \# basic blocks with a single predecessor and two successors&ft37& \#  times the address of a variables is taken ("\&" in C)\\ \hline
		ft10& \# basic blocks with a two predecessors and one successor&ft38& \#  times the address of a function is taken ("\&" in C)\\ \hline
		ft11& \# basic blocks with two successors and two predecessors& ft39& \#  indirect calls (i.e. done via pointers) in the method\\ \hline
		ft12& \# basic blocks with more then two successors and more then two&ft40& \#  assignment instructions with the left operand an integer constant in the method\\ \hline
		ft13& \# basic blocks with \#  instructions less then 15&ft41& \#  binary operations with one of the operands an integer constant in the method\\ \hline
		ft14& \# basic blocks with \#  instructions in the interval [15, 500]&ft42& \#  calls with pointers as arguments\\ \hline
		ft15& \# basic blocks with \#  instructions greater then 500&ft43& \#  calls with the \#  arguments is greater then 4\\ \hline
		ft16& \# edges in the control flow graph &ft44& \#  calls that return a pointer\\ \hline
		ft17& \# critical edges in the control flow graph&ft45& \#  calls that return an integer\\ \hline
		ft18& \# abnormal edges in the control flow graph&ft46& \#  occurrences of integer constant zero\\ \hline
		ft19& \# direct calls in the method&ft47& \#  occurrences of 32-bit integer constants\\ \hline
		ft20& \# conditional branches in the method&ft48& \#  occurrences of integer constant one\\ \hline
		ft21& \# assignment instructions in the method&ft49& \#  occurrences of 64-bit integer constants\\ \hline
		ft22& \# binary integer operations in the method&ft50& \#  references of a local variables in the method\\ \hline
		ft23& \# binary floating point operations in the method&ft51& \#  references (def/use) of static/extern variables in the method\\ \hline
		ft24& \#instructions in the method&ft52& \#  local variables referred in the method\\ \hline
		ft25& Average of \#  instructions in basic blocks&ft53& \#  static/extern variables referred in the method\\ \hline
		ft26& Average of \#  phi-nodes at the beginning of a basic block&ft54& \#  local variables that are pointers in the method\\ \hline
		ft27& Average of arguments for a phi-node predecessors&ft55& \#  static/extern variables that are pointers in the method\\ \hline
		ft28& \# basic blocks with no phi nodes&ft56& \#  unconditional branches in the method\\ \hline
	\end{tabular}
	\vspace{-0.4cm}
\end{table}
\linespread{1}

Program characterization, a process to numerically or structurally quantify a target program, allows modern machine-learning techniques to be applied to compiler tasks, since most machine-learning methods assume numerical inputs for both training and testing data. Program characterization is usually achieved by extracting from the target program a set of \textit{static} features and/or a set of \textit{dyanamic} features. Static features can be obtained directly from the source code or intermediate representation of the target program, while the procurement of dynamic features usually requires actually executing the target program to capture certain run-time behavior.

With current intensive research on program characterization, new features, both static and dynamic, are continuously being proposed. An example of a static feature set is shown in Table \ref{table_56_features}, which lists all the 56 original static features extracted by Milpost GCC from cTuning Compiler Collection \cite{staticfeatures}, with many of them being different yet non-independent features. For example. ft. 8 implies ft. 2 and ft. 5, meaning that these features are correlated.
In a compiler task, determining which features to use or to include for program characterization is of considerable importance, since different features can have different effects, and hence different resulting performance on a specific compiler task. Intuitively, including as many features as possible for program characterization seems to be a reasonable approach when considering feature selection. This, however, essentially increases the dimensionality of the feature space, and thus potentially introduces extra computational overhead. Also, some features may not be relevant to the target compiler task, and therefore behave equivalently as noise in program characterization, harming the resulting performance. Furthermore, many features, though different, capture very similar characteristics of a program. The similarities among features can produce bias that overemphasizes certain aspects of a program's characteristics, and consequently lead to an inaccurate result for a target compiler task. Due to the aforementioned reasons, many machine-learning-aided compiler tasks still heavily rely on expert knowledge for determining an appropriate set of features to use. This, unfortunately, hinders the full automation of compiler tasks using machine learning, which is originally deemed as a tool to lower the involvement of field expertise and engineer intervention.

\begin{figure*}[t]
	\centering
	\includegraphics[width=7in]{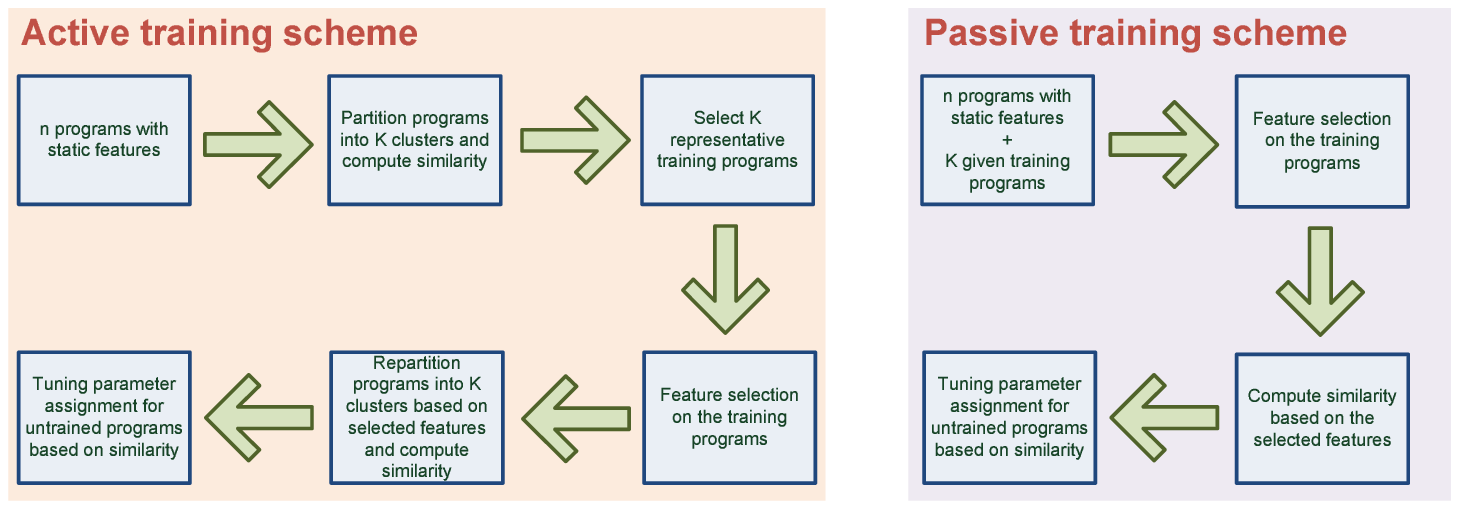}
	\caption{Flow diagrams of the proposed methods.}
	\label{Fig_CGO}
	\vspace{-4mm}
\end{figure*}

This paper proposes \textit{FEAture Selection for compiler Tasks} (\textit{FEAST}), an automated framework of feature selection for a target compiler task. FEAST incorporates into its framework a variety of feature selection methods, including the well-known LASSO \cite{zou2005regularization}, sequential forward and backward selection. Given a compiler task and a list of feature candidates, FEAST first samples a small set of available programs as training data, and then uses the integrated feature selection methods to choose $M$ most appropriate or relevant features for this specific compiler tasks. The remaining programs can then be handled using only the chosen features.

To demonstrate the feasibility and practicality of FEAST, we assess its performance on a proposed method to assignment of compiler parameters. Modern compilers usually embed a rich set of tuning parameters to provide hints and guidance for their optimization processes. To obtain an optimal compiled executable program, exhaustive trials over all combinations of tuning parameters of the utilized compiler is required. This is, in general, excessively time consuming and hence infeasible due to its combinatorial nature. As many other compiler tasks, in practice, expert intervention is frequently triggered and heuristics based on expertise and experience are adopted as a general guideline for tuning parameters \cite{agakov2006using}. Unfortunately, fine tuning compiler's parameters may require multiple compilation processes, and can take up to weeks or months to complete for a moderate program size, entailing a huge burden and software engineering. In this work, as a test case for FEAST, we develop an automated method to assigning ``good'' parameters to a pool of programs using machine-learning techniques. We then show that using FEAST, the dimension of the feature space can be greatly reduced to pertaining relevant features, while maintaining comparable resulting performance.

This paper is organized as follows. Sec. \ref{FEAST and Compiler Parameter Assignment} details the proposed FEAST framework as well as the automatic compiler parameter assignment process. Sec. \ref{Performance Evaluation} presents experimental results with detailed analysis. Related work is discussed in Sec. \ref{Related Work}. Finally, Sec. \ref{Conclusions} draws some conclusions and provides envisions for future work.

\section{FEAST and Compiler Parameter Assignment}
\label{FEAST and Compiler Parameter Assignment}

This section details the mechanisms of FEAST, and presents the proposed compiler parameter assignment method.

\subsection{FEAST}
\label{subsec_FEAST}

Given $K$ training programs and a set of $p$ numerical features, FEAST assumes a linear model for resulting performance and feature values:
\begin{equation}
y = \textnormal{X}\beta + \beta_0\textbf{1}
\end{equation}
where $y \in \mathbb{R}^{K\times 1}$ is the compiled programs' performance vector, whose $i$-th entry denotes the performance (measured in some pre-defined metrics) of the $i$-th program in the set of total $K$ training programs. $\textnormal{X} \in \mathbb{R}^{K\times p}$ is a matrix whose ($i,j$)-th entry denotes the value of the $j$-th extracted feature of the $i$-th program. $\beta \in \mathbb{R}^{p\times 1}$ and $\beta_0\in \mathbb{R}$ are the coefficients describing the linear relationship between the features and the resulting performance. \textbf{1} is a ${K\times 1}$ vector with its elements being all 1s.
FEAST utilizes three widely used feature selection methods, namely LASSO, sequential forward selection (SFS) and sequential backward selection (SBS), all of which pick $M$ most influential features out of the total $p$ features. Specifically, the elastic net approach for LASSO adopted in FEAST selects features by first solving optimization problem \cite{zou2005regularization}:
\begin{equation}
\textnormal{min}_{\tilde{\beta}} \  \frac{1}{K} \left\| y - \tilde{X}\tilde{\beta} \right\|^{2}_{2} + \lambda\left\| \tilde{\beta} \right\|_{1}
\end{equation}
where $ \tilde{X} = \big[ X \ \textbf{1}\big]$. The first $p$ elements of the solution $\tilde{\beta}$ are the coefficient estimates whose magnitudes directly reflect the influence of the corresponding features. $M$ selected features are chosen as those with coefficients of largest magnitudes. SFS and SBS are other well-known feature selection methods. For SFS, we sequentially or greedily select the most relevant feature from the training programs until we have selected $M$ features. For SBS,  we sequentially exclude the most irrelevant feature until there are $M$ features left.  We omit algorithmic descriptions of SFS and SBS, and refer interested readers to \cite{dash1997feature} for implementation details.

\subsection{Compiler Parameter Assignment Algorithms}
This section discusses the proposed method for compiler parameter assignment algorithm and the application of FEAST to this task. 

Given a compiler with many parameters to set, finding an optimal assignment of parameters that can best compile a target program is a vexing problem. This paper proposes a machine-learning algorithm that automatically assigns ``good'' parameters to target programs based on the known optimal assignment to the training programs. The proposed method works in two schemes: \textit{active training scheme} and \textit{passive training scheme} (see Fig. \ref{Fig_CGO}). In active training scheme, the users can actively acquire the optimal compiler parameters for a subset of programs, while in passive training scheme, the users are given as prior knowledge a set of programs whose optimal compiler parameters are known. A practical example of active training scheme is that a company has no prior knowledge about compiler parameter assignment, and it has a very limited budget which only allows it to select a small set of programs for full tuning. Remaining large set of programs, however, has to be quickly and efficiently compiled. For passive training scheme, a company has a small set of programs with well-tuned compiler parameters, and it would like to quickly find good compiler parameters for other programs. Note that, in the active training scheme, our proposed method can also automatically choose a good set of candidate programs for full tuning. The remaining section is dedicated to detailing the two preceding schemes with FEAST application to them respectively.

\textbf{Active training scheme.}
Since acquirement of dynamic features can be very expensive due to the potential need for multiple compilations and iterative tuning, we opt to use numerical static features in the proposed compiler parameter assignment task. Also, we use the execution time of a compiled program as the measure of performance for that program. Given $n$ programs with $p$ static features, we are granted to choose $K \leq n$ programs as training samples and acquire the optimal compiler parameters for each training program that optimize the execution time. In order to choose the $K$ programs, we first compute the similarity of each program pair based on the Euclidean distance between the corresponding static feature vectors, and partition the programs into $K$ clusters using K-means clustering. For each cluster, we select one program that has the least average distance to other programs in the same cluster as a training program. Exhaustive trials are then conducted for the training programs to obtain their optimal compiler parameters and the associated execution time. Given the $K$ training programs as well as their execution time, FEAST can then select $M$ features that are most influential to the training programs' performance (execution time in this case) . We then leverage the selected features to recompute similarities and repartition the programs. Lastly, each untrained program is assigned by the optimal parameters of the most similar training program. See Alg. \ref{algo_active} for detailed algorithm.

\begin{algorithm}[!t]
	\caption{Active Training Scheme}
	\label{algo_active}
	\begin{algorithmic}	
		\State \textbf{Input:} $n$ programs with $p$ static features, number $K$ of training programs for optimization
		\State \textbf{Output:} compiler parameter assignment for each untrained program
		\State \textbf{Procedures:}
		\begin{enumerate}
			\item Partition $n$ programs into $K$ clusters by K-means clustering
			\item For each cluster select one program having the least sum of distances to other programs in the same cluster as a training program.
			\item Find the set of optimal compiler parameters of the selected $K$ programs.
			\item Use FEAST to perform feature selection on the selected $K$ programs by regressing their optimal execution time with respect to their static features.
			\item Repartition the untrained programs based on the similarities computed by the selected features.
			\item For each untrained program, assign the compiler parameters of the closest trained program to it based on the selected features.
		\end{enumerate}
	\end{algorithmic}
\end{algorithm}

\begin{algorithm}
	\caption{Passive Training Scheme}
	\label{algo_passive}
	\begin{algorithmic}	
		\State \textbf{Input:} $n$ programs with $p$ static features, $K$ given training programs
		\State \textbf{Output:} compiler parameter assignment for each untrained program
		\State \textbf{Procedures:}
		\begin{enumerate}
			\item Find the set of optimal compiler parameters of the  $K$ given training programs.
			\item Use FEAST to perform feature selection on the selected $K$ programs by regressing their optimal execution time with respect to their static features.
			\item Compute distances between programs based on the selected features.
			\item For each untrained program, assign the compiler parameters of the closest trained program based on the selected features.
		\end{enumerate}
	\end{algorithmic}
\end{algorithm}

\textbf{Passive training scheme.} Different from the active training scheme, in addition to $n$ programs with $p$ static features, we are also given $K$ pre-selected training programs. Therefore, the methodology of the passive training scheme is similar to that of the active training scheme, except that the clustering procedures described in active training scheme are no longer required. See Alg. \ref{algo_passive} for detailed algorithm.

\section{Performance Evaluation of Complier Parameter Assignment}
\label{Performance Evaluation}
We test our implementations using the PolyBench benchmark suite \cite{Data} that consists of $n=$ 30 programs. The programs are characterized using $p=56$ static features from cTuning Compiler Collection \cite{staticfeatures}. For the two proposed training schemes, $K$ trained programs are used for feature selection and compiler parameter assignment, and for each feature selection method (LASSO, SFS, SBS), we select the $M=10$ most relevant features.

\subsection{Performance Comparison of the Active and Passive Training Schemes}

\begin{figure*}[t]
	\begin{subfigure}[b]{0.5\textwidth}
		\includegraphics[width=\linewidth]{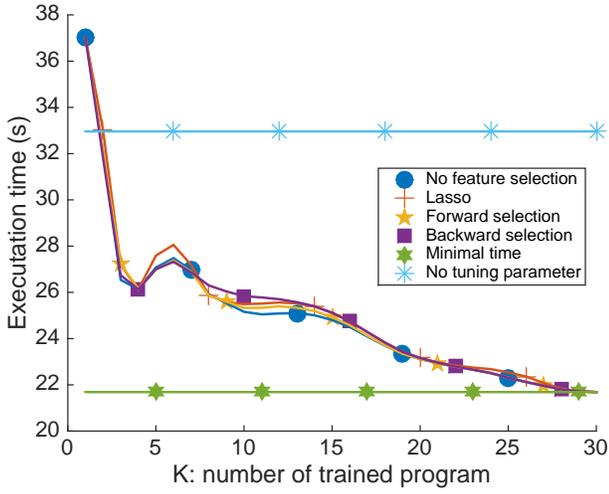}
		\caption{Active training scheme.}
	\end{subfigure}%
	\begin{subfigure}[b]{0.5\textwidth}
		\includegraphics[width=\linewidth]{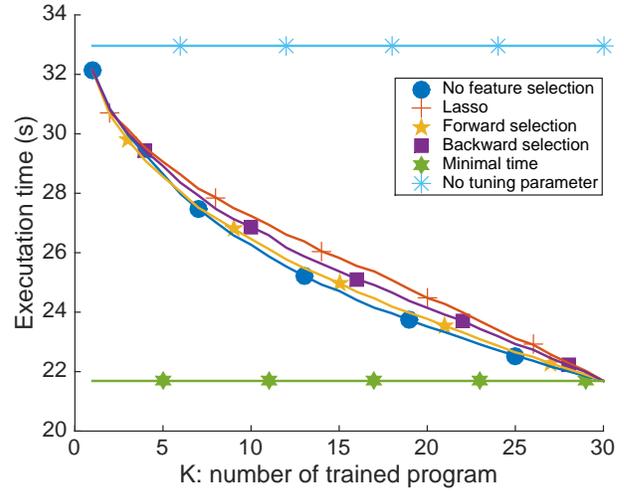}
		\caption{Passive training scheme.}
	\end{subfigure}%
	\caption{Program execution time of the active and passive training schemes. Minimal time refers to exhaustive parameter optimization on every program. It can be observed that the execution time of the three feature selection algorithms integrated in FEAST are comparable to that of using all features, which strongly suggests that important features affecting program execution are indeed identified by FEAST.}
	\label{Fig_time}
\end{figure*}
\begin{figure*}[!t]
	\centering
	\includegraphics[width=7.5in]{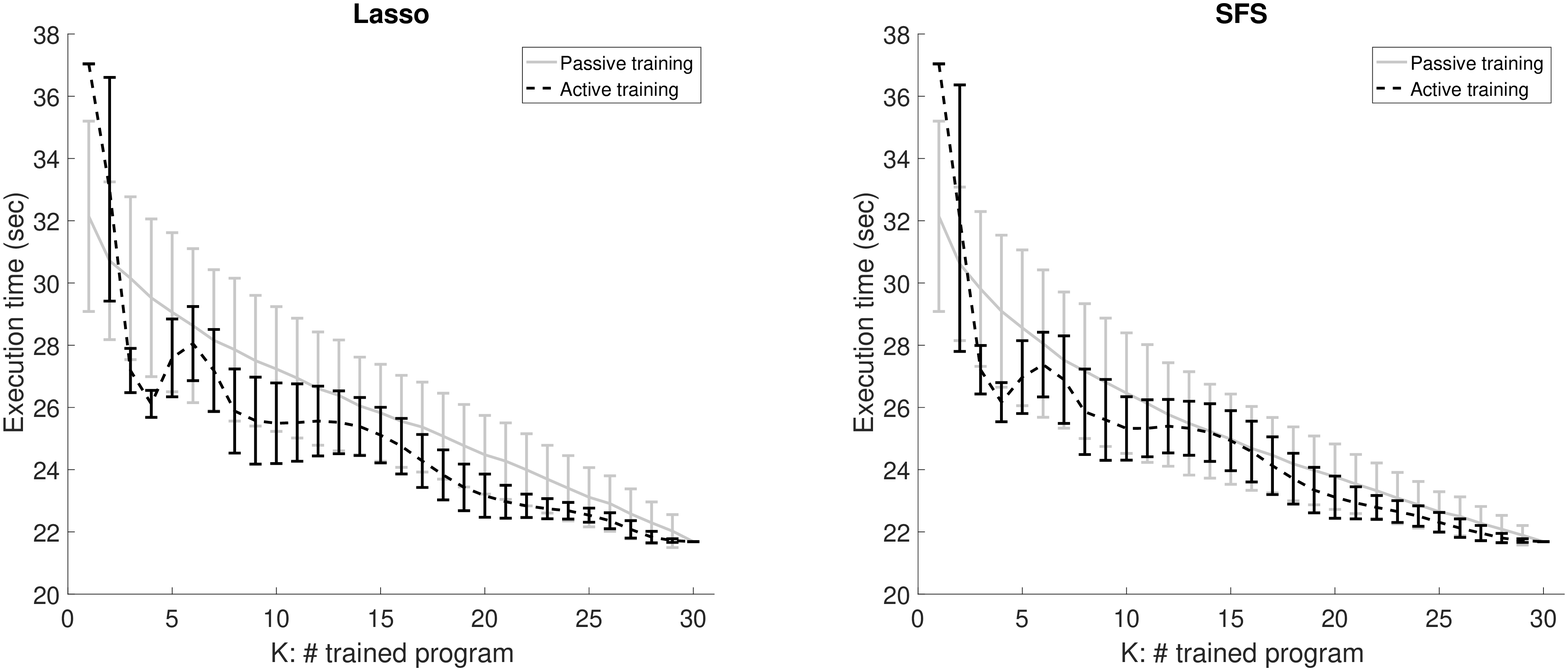}
	\caption{Comparison of active and passive training schemes. Error bars represent standard deviation. The average execution time of active training is smaller that that of passive training. The variations in active training scheme are caused by random initialization in K-means clustering procedure, whereas the variations in passive training scheme are caused by randomness in the selection of training programs. }
	\label{Fig_error_bar}
\end{figure*}

\begin{figure*}[!t]
	\begin{subfigure}[b]{0.5\textwidth}
		\includegraphics[width=\linewidth]{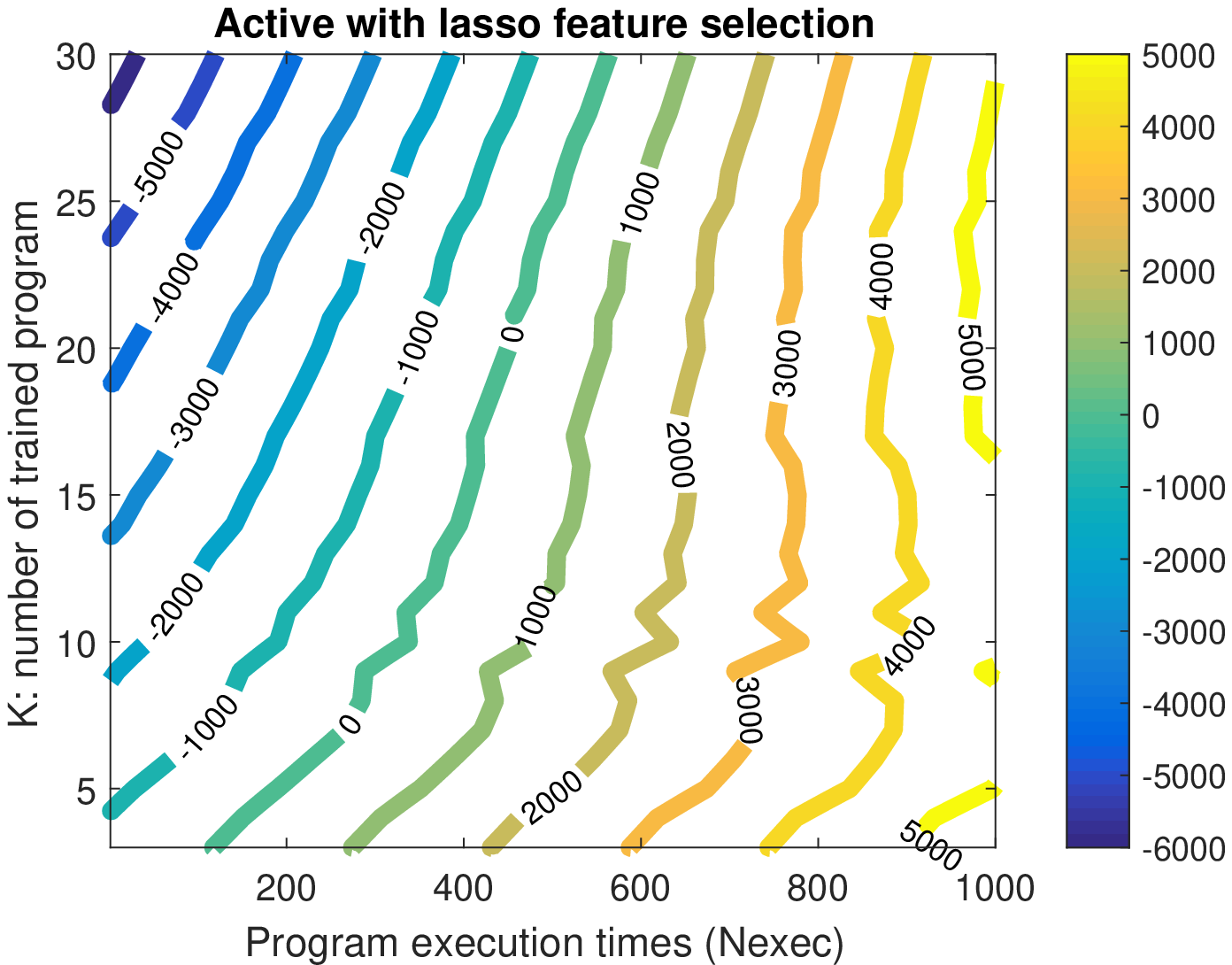}
		\caption{Active training scheme.}
	\end{subfigure}%
	\begin{subfigure}[b]{0.5\textwidth}
		\includegraphics[width=\linewidth]{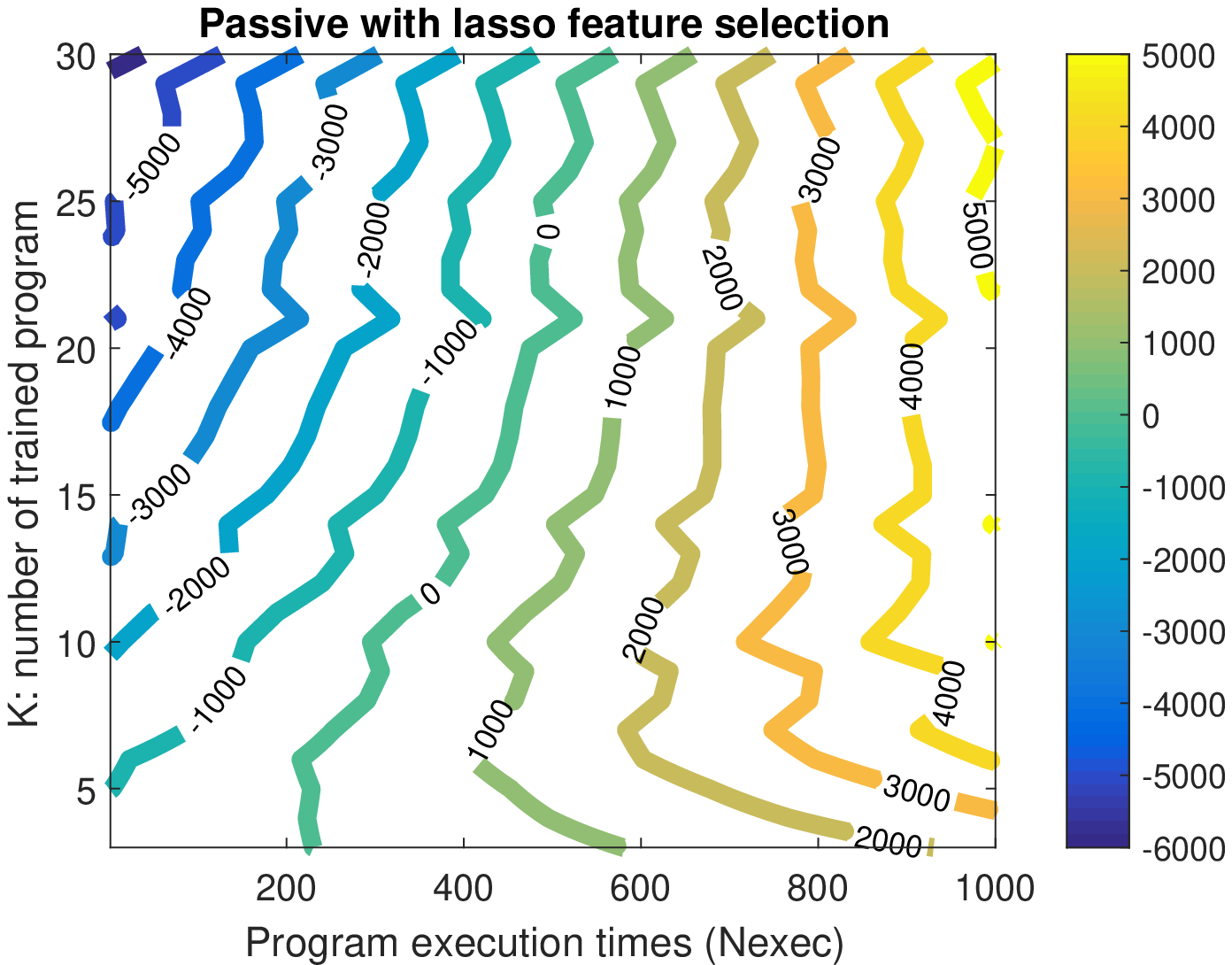}
		\caption{Passive training scheme.}
	\end{subfigure}%
	\caption{Time reduction for active and passive training schemes using LASSO feature selection. The parameter \textnormal{Nexec} specifies the number of times a program is executed. The contour indicates a phase transition in time reduction. The figure suggests that with Nexec large enough, the proposed compiler parameter assignment method can provide time saving when considering overall execution time.}
	\label{Fig_coutour}
\end{figure*}

Fig. \ref{Fig_time} shows the program execution time of active and passive training schemes. The minimal execution time refers to the optimal performance over 192 possible combinations of compiler parameters of every program. We also show the results for the case where no tunable compiler parameters are enabled. The results of minimal-time and no-tuning-parameter
are regarded as baselines for comparing the proposed FEAST methods, as their execution time does not depend on the number $K$ of training programs. Furthermore, to validate the utility of FEAST, we also calculate the execution time using all features, i.e., the case where FEAST is disabled. 

For the three feature selection methods introduced in Sec. \ref{subsec_FEAST} (LASSO, SFS, and SBS), we select the top $M = 10$ important features and use these selected features to compute program similarities and assign compiler parameters. In this setting, we only compare the execution time of the untrained programs, and we omit the computation time required to obtain the optimal compiler parameters associated with the training programs. We will consider the overall execution time of the trained and untrained programs shortly.

In Fig. \ref{Fig_time}, each untrained program is executed once and we sum up the execution time to get the overall execution time under various values of $K$ of trained programs. 
It is observed that the execution time of both training schemes converge to the minimal execution time as $K$ increases. 
The curve of the passive training scheme is smoother than that of the active training scheme due to the fact that the former is an averaged result over 1000 trials of randomly selected training programs.

The trends in Fig. \ref{Fig_time} can be explained as follows: when adopting passive training, every program has an equal chance to be selected as a training program. As $K$ grows, the set of available optimal compiler parameters for training programs increases as well, resulting in the decrease in average total execution time. Active training, on the other hand, adopts K-means clustering when selecting the training programs. Given a fixed $K$, only $K$ programs can be selected for training and compiler parameter optimization. Therefore, for small $K$ (e.g.,. $K = 1$ or $2$), we might select the programs whose optimal compiler parameters do not fully benefit other untrained programs.

The execution time of cases with FEAST enabled are shown to be comparable to that of using all features, which strongly suggests that FEAST can successfully select important features affecting program execution, leading to dimensionality reduction while still attaining satisfactory execution time reduction.

To gain more insights on the performance of active and passive training schemes, we compare the execution time of LASSO and SFS in Fig. \ref{Fig_error_bar}. The comparison of SBS is omitted since in practice SBS is computationally inefficient due to its sequential feature removal nature, especially for high-dimensional data. It is observed in Fig. \ref{Fig_error_bar} that the average execution time of active training is smaller than that of passive training, since for active training, we are able to select representative programs as training samples for optimization. These results indicate that the execution time can be further reduced when we have the freedom to select $K$ representative programs from clustering as training samples for compiler parameter assignment.

\subsection{Overall Execution Time Comparison}
In this section, we consider the overall execution time, including time overhead imposed by the proposed algorithms as well as the execution time of every untrained program. To this end, we introduce a parameter \textbf{Nexec}, the number of executions per program. The motivation behind introducing \textnormal{Nexec} is as follows: for programs such as matrix operations, users may execute these core programs multiple times (i.e., \textnormal{Nexec} times). Therefore, as will be demonstrated by the analysis detailed in this section, the time overhead introduced by the proposed algorithms will be compensated as \textnormal{Nexec} increases, since the time spent in optimizing the training programs comprises relatively small portion of the overall time cost with large Nexec. \textbf{Time Reduction (TR)} can therefore be computed using the following formula:
\begin{align}
\textnormal{TR} = \textnormal{Nexec} \cdot T_{\text{null}} -
\textnormal{Nexec} \cdot T_{\text{auto}} -
T^K_{\text{exhaustive}},  \nonumber
\end{align}
where \textnormal{Nexec} denotes the number of executions for each program, $T_{\text{null}}$ represents the total time to run every program with all tunable compiler parameters disabled, $T_{\text{auto}}$ denotes the total time to run every program compiled by using the proposed compiler parameter assignment method, and $T^K_{\text{exhaustive}}$ denotes the computation time for finding the optimal compiler parameters for $K$ training programs.

Fig. \ref{Fig_coutour} shows the contour plot of the overall time reduction metric for both training schemes with LASSO.
Results using SFS and SBS are similar, and are omitted in this paper. If TR is positive, it indicates the overall execution time is smaller than that with all tunable compiler parameters disabled. It can be observed that for each $K$, TR becomes positive when \textnormal{Nexec} exceeds certain threshold value, implying that if a user uses the compiled programs repeatedly, the proposed method could potentially provide great time saving. Also for each $K$, the preceding threshold of active training is lower than that of passive training, since programs to exhaustively optimize are actively chosen by the proposed method.

\begin{table*}[!h]
	\caption{Top 10 selected features from various methods integrated in FEAST. The number in the bracket indicates the feature ranking for each method.}
	\label{table_feature}
	\centering
	\begin{tabular}{l|l|l}
		\hline
		LASSO&SFS&SBS                                                                                                                                            \\ \hline
		\begin{tabular}[c]{@{}l@{}}Number of basic blocks\\ with a single predecessor \\ and a single successor (6)\end{tabular}                        & \begin{tabular}[c]{@{}l@{}}Number of basic blocks in \\ the method (3)\end{tabular}                                                               & \begin{tabular}[c]{@{}l@{}}Number of basic blocks \\ with a two predecessors \\ and one successor (8)\end{tabular}                             \\ \hline
		\begin{tabular}[c]{@{}l@{}}Number of basic blocks \\ with a single predecessor \\ and two successors (7)\end{tabular}                           & \begin{tabular}[c]{@{}l@{}}Number of basic blocks with\\ a two predecessors and one \\ successor (7)\end{tabular}                                 & \begin{tabular}[c]{@{}l@{}}Number of conditional \\ branches in the method (5)\end{tabular}                                                    \\ \hline
		\begin{tabular}[c]{@{}l@{}}Number of basic blocks \\ with more then two \\ successors and more than \\ two predecessors  (8)\end{tabular}       & \begin{tabular}[c]{@{}l@{}}Number of basic blocks with \\ two successors and two \\ predecessors (2)\end{tabular}                                 & \begin{tabular}[c]{@{}l@{}}Number of instructions \\ in the method (9)\end{tabular}                                                            \\ \hline
		\begin{tabular}[c]{@{}l@{}}Number of basic blocks \\ with number of instructions \\ in the interval {[}15, 500{]} (5)\end{tabular}              & \begin{tabular}[c]{@{}l@{}}Number of basic blocks with \\ more then two successors and \\ more than two predecessors (6)\end{tabular}             & \begin{tabular}[c]{@{}l@{}}Average of number of \\ phi-nodes at the beginning \\ of a basic block (10)\end{tabular}                            \\ \hline
		\begin{tabular}[c]{@{}l@{}}Number of assignment \\ instructions in the \\ method (9)\end{tabular}                                               & \begin{tabular}[c]{@{}l@{}}Number of basic blocks with \\ number of instructions greater \\ then 500 (4)\end{tabular}                             & \begin{tabular}[c]{@{}l@{}}Number of basic blocks \\ with more than 3 phi \\ nodes (4)\end{tabular}                                            \\ \hline
		\begin{tabular}[c]{@{}l@{}}Number of binary integer \\ operations in the method (1)\end{tabular}                                                & \begin{tabular}[c]{@{}l@{}}Number of direct calls in the \\ method (9)\end{tabular}                                                               & \begin{tabular}[c]{@{}l@{}}Number of basic block \\ where total number of \\ arguments for all phi-nodes \\ is greater than 5 (7)\end{tabular} \\ \hline
		\begin{tabular}[c]{@{}l@{}}Number of binary floating \\ point operations in the \\ method (2)\end{tabular}                                      & \begin{tabular}[c]{@{}l@{}}Number of assignment \\ instructions in the \\ method (10)\end{tabular}                                                & \begin{tabular}[c]{@{}l@{}}Number of switch\\ instructions in the \\ method (3)\end{tabular}                                                   \\ \hline
		\begin{tabular}[c]{@{}l@{}}Number of basic blocks \\ with phi nodes in the \\ interval {[}0,3{]} (4)\end{tabular}                               & \begin{tabular}[c]{@{}l@{}}Number of binary integer \\ operations in the method  (1)\end{tabular}                                                 & \begin{tabular}[c]{@{}l@{}}Number of unary operations \\ in the method (2)\end{tabular}                                                        \\ \hline
		\begin{tabular}[c]{@{}l@{}}Number of basic block \\ where total number of \\ arguments for all phi-nodes \\ is greater than 5 (10)\end{tabular} & \begin{tabular}[c]{@{}l@{}}Number of basic blocks with \\ more than 3 phi nodes  (5)\end{tabular}                                                 & \begin{tabular}[c]{@{}l@{}}Number of assignment \\ instructions with the left \\ operand an integer constant \\ in the method (6)\end{tabular} \\ \hline
		\begin{tabular}[c]{@{}l@{}}Number of unary operations \\ in the method (3)\end{tabular}                                                         & \begin{tabular}[c]{@{}l@{}}Number of basic block \\ where total number of \\ arguments for all phi-nodes \\ is in greater than 5 (8)\end{tabular} & \begin{tabular}[c]{@{}l@{}}Number of binary operations\\ with one of the operands an \\ integer constant in the \\ method (1)\end{tabular}     \\ \hline
	\end{tabular}
\end{table*}

\subsection{Features Selected by FEAST}

To investigate the key features affecting compiler execution time, we inspect the selected features from FEAST.
Table \ref{table_feature} lists the top 10 selected features using various feature selection methods integrated in FEAST. The features are selected with 3-fold cross validation method for regressing the optimal execution time of all 30 programs with respect to their static program features. Based on the selected features,  we can categorize the selected features into three groups: 1) Control Flow Graph (CFG), 2) assignments/operations, and 3) phi node. CFG features describe a programs' control flow, which can be largely influenced by instruction branches, such as ``if-else'',  ``if- if'' and ``for loop'' statements. The selected CFG features are reasonable as in our program testing dataset, “for loop” contributes to the major part of programs' control flow. In addition, assignment operations are essential to matrix operations, and hence possess discriminative power for distinguishing programs. Lastly, Phi node is a special operation of Intermediate Representation (IR). It is designed for Static Single Assignment form (SSA) which enables a compiler to perform further optimization, and hence it is an important factor for program execution.


\section{Related Work}
\label{Related Work}
Recent rapid development of \textit{program characterization}, a process to quantify programs, allows the application of modern machine-learning techniques to the field fo code compilation and optimization. These machine-learning techniques provide powerful tools that are widely used in various aspects of compilation procedures. For example, Buse and Weimer use static features to identify "hot paths" (executional paths that are most frequently invoked) of a target program by applying logistic regression \cite{buse2009road} without ever profiling the program, Kulkarni et al. build an evolving neural-network model that uses static features to help guide the inlining heuristics in compilation process \cite{kulkarni2013automatic}, and Wang and O'Boyle exploit the use of artificial neural networks (ANNs) as well as support vector machine (SVM) for automatic code parallelization on multi-core systems \cite{wang2009mapping}, among others. Many existing applications of machine-learning-enabled compiler tasks use features, either static or dynamic, selected by the designers, and hence heavily rely on field expertise. This work provides a comprehensive solution to this problem by using modern statistical methods to select appropriate features for a specific case.

There has also been a vast amount of research dedicated to designing suitable features for target applications. For example, it is shown in \cite{demme2012approximate} that, for compiler tasks that cannot afford the time cost for procurement of dynamic features, carefully designed graph-based static features can achieve accuracy and performance comparable to dynamic features in some applications, regardless the fact that it is prevailingly believed dynamic features are preferred due to insightful information they provide. Another example is the compiler parameters assignment task by Park et al. \cite{park2012using}. In their work, an SVM-based supervised training algorithm is used to train a set of support vectors that can help estimate the performance or reaction of an unknown program to a set of compiler parameters. They use newly-defined graph-based static features for training, which achieves high performance comparable to that using dynamic features, but without the need to invoke multiple compilations and profiling. While in general, it is possible to design dedicated features for specific tasks, the applicability of these dedicated features to other applications remain questionable. In the scenario where there are excessive number of numerical features that may or may not fit a target task, FEAST can help in selecting the most meaningful and influential features.

As to the task of compiler parameter tuning, recent work has been focused on automation of parameter assignment process. Compiler parameter tuning has long been a crucial problem which attracts a vast amount of attention. Recent trends and efforts on exploiting the power of modern machine-learning techniques have achieved tremendous success in compiler parameter tuning. Stephenson and Amarasinghe demonstrate the potential of machine learning on automatic compiler parameter tuning by applying ANNs and SVM to predict a suitable loop-unrolling factor for a target program \cite{stephenson2005predicting}. This work is relatively restricted, since it deals with a single compiler parameter. Agakov et al. propose a computer-aided compiler parameter tuning method by identifying similar programs using static features \cite{agakov2006using}, where certain level of expert intervention is still required. Cavazos et al. characterize programs with dynamic features, and use logistic regression, a classic example of conventional machine-learning algorithm, to predict good compiler parameter assignment \cite{cavazos2007rapidly}. While providing state-of-the-art performance, \cite{cavazos2007rapidly} requires dynamic features, which can be expensive to acquire. In \cite{park2012using}, graph-based features are used along with SVM for performance prediction given a compilation sequence. This work uses dedicated features for the machine-learning task, and further implicitly utilizes excessive number of candidate assignments of compiler parameters in order to find a good assignment, resulting in a non-scalable algorithm. On the other hand, the proposed compiler parameter assignment algorithm is a comprehensive assignment algorithm that does not require dynamic features or dedicated static features. Furthermore, the training data that need full optimization can be set fixed. The good assignment for an unseen target program is directly derived from that of trained programs, implying its scalability with the number of potential assignments.

\section{Conclusions and Future Work}
\label{Conclusions}
In this work, we propose FEAST, an automated framework for feature selection in compiler tasks that incorporates with well-known feature selection methods including LASSO, sequential forward and backward selection. We demonstrate the feasibility and applicability of FEAST by testing it on a proposed method for the task of compiler parameter assignment. The experimental results show that the three feature selection methods integrated in FEAST can select a representative small subset of static features that, when used in the compiler parameter assignment task, can achieve non-compromised performance. We also validate the effectiveness of the proposed methods by experimentally demonstrating significant overall execution time reduction of our method in a practical scenario where each program is required to run multiple times. Lastly, we discussed the roles of the features selected by FEAST, which provides deep insights into compilation procedures.
In summary, our contributions are two-fold:
\begin{enumerate}
	\item We integrate into FEAST with various modern machine-learning and optimization techniques for feature selection for compilation tasks.
	\item We demonstrate the applicability of FEAST by experimentally showing that it can achieve comparable performance in compiler parameter assignment tasks with a very small set of selected static features.
\end{enumerate}

For future work, we are interested in exploring the inherent structural dependencies of codes in each program as additional features for compiler parameter assignment. We are also interested in integrating the proposed compiler parameter assignment algorithm with recently developed automated community detection algorithms, such as AMOS \cite{chen2016amos}, to automatically cluster similar programs for the proposed passive and active training schemes.

\small{
	
	\bibliographystyle{IEEEtran}
	\bibliography{IEEEabrv,CGO_bib}

\begin{thebibliography}{10}
\providecommand{\url}[1]{#1}
\csname url@samestyle\endcsname
\providecommand{\newblock}{\relax}
\providecommand{\bibinfo}[2]{#2}
\providecommand{\BIBentrySTDinterwordspacing}{\spaceskip=0pt\relax}
\providecommand{\BIBentryALTinterwordstretchfactor}{4}
\providecommand{\BIBentryALTinterwordspacing}{\spaceskip=\fontdimen2\font plus
\BIBentryALTinterwordstretchfactor\fontdimen3\font minus
  \fontdimen4\font\relax}
\providecommand{\BIBforeignlanguage}[2]{{%
\expandafter\ifx\csname l@#1\endcsname\relax
\typeout{** WARNING: IEEEtran.bst: No hyphenation pattern has been}%
\typeout{** loaded for the language `#1'. Using the pattern for}%
\typeout{** the default language instead.}%
\else
\language=\csname l@#1\endcsname
\fi
#2}}
\providecommand{\BIBdecl}{\relax}
\BIBdecl

\bibitem{staticfeatures}
\BIBentryALTinterwordspacing
``c{T}uning {C}ompiler {C}ollection.'' [Online]. Available:
  \url{http://ctuning.org/wiki/index.php?title=CTools:CTuningCC}
\BIBentrySTDinterwordspacing

\bibitem{zou2005regularization}
H.~Zou and T.~Hastie, ``Regularization and variable selection via the elastic
  net,'' \emph{Journal of the Royal Statistical Society: Series B (Statistical
  Methodology)}, vol.~67, no.~2, pp. 301--320, 2005.

\bibitem{agakov2006using}
F.~Agakov, E.~Bonilla, J.~Cavazos, B.~Franke, G.~Fursin, M.~F. O'Boyle,
  J.~Thomson, M.~Toussaint, and C.~K. Williams, ``Using machine learning to
  focus iterative optimization,'' in \emph{Proceedings of the International
  Symposium on Code Generation and Optimization}.\hskip 1em plus 0.5em minus
  0.4em\relax IEEE Computer Society, 2006, pp. 295--305.

\bibitem{dash1997feature}
M.~Dash and H.~Liu, ``Feature selection for classification,'' \emph{Intelligent
  data analysis}, vol.~1, no.~1, pp. 131--156, 1997.

\bibitem{Data}
\BIBentryALTinterwordspacing
``Polybench benchmark suite.'' [Online]. Available:
  \url{http://web.cse.ohiostate.edu/~pouchet/software/polybench/}
\BIBentrySTDinterwordspacing

\bibitem{buse2009road}
R.~P. Buse and W.~Weimer, ``The road not taken: Estimating path execution
  frequency statically,'' in \emph{Proceedings of the 31st International
  Conference on Software Engineering}.\hskip 1em plus 0.5em minus 0.4em\relax
  IEEE Computer Society, 2009, pp. 144--154.

\bibitem{kulkarni2013automatic}
S.~Kulkarni, J.~Cavazos, C.~Wimmer, and D.~Simon, ``Automatic construction of
  inlining heuristics using machine learning,'' in \emph{Code Generation and
  Optimization (CGO), 2013 IEEE/ACM International Symposium on}.\hskip 1em plus
  0.5em minus 0.4em\relax IEEE, 2013, pp. 1--12.

\bibitem{wang2009mapping}
Z.~Wang and M.~F. O'Boyle, ``Mapping parallelism to multi-cores: a machine
  learning based approach,'' in \emph{ACM Sigplan notices}, vol.~44,
  no.~4.\hskip 1em plus 0.5em minus 0.4em\relax ACM, 2009, pp. 75--84.

\bibitem{demme2012approximate}
J.~Demme and S.~Sethumadhavan, ``Approximate graph clustering for program
  characterization,'' \emph{ACM Transactions on Architecture and Code
  Optimization (TACO)}, vol.~8, no.~4, p.~21, 2012.

\bibitem{stephenson2005predicting}
M.~Stephenson and S.~Amarasinghe, ``Predicting unroll factors using supervised
  classification,'' in \emph{International symposium on code generation and
  optimization}.\hskip 1em plus 0.5em minus 0.4em\relax IEEE, 2005, pp.
  123--134.

\bibitem{cavazos2007rapidly}
J.~Cavazos, G.~Fursin, F.~Agakov, E.~Bonilla, M.~F. O'Boyle, and O.~Temam,
  ``Rapidly selecting good compiler optimizations using performance counters,''
  in \emph{International Symposium on Code Generation and Optimization
  (CGO'07)}.\hskip 1em plus 0.5em minus 0.4em\relax IEEE, 2007, pp. 185--197.

\bibitem{park2012using}
E.~Park, J.~Cavazos, and M.~A. Alvarez, ``Using graph-based program
  characterization for predictive modeling,'' in \emph{Proceedings of the Tenth
  International Symposium on Code Generation and Optimization}.\hskip 1em plus
  0.5em minus 0.4em\relax ACM, 2012, pp. 196--206.

\bibitem{chen2016amos}
P.-Y. Chen, T.~Gensollen, and A.~O. Hero~III, ``Amos: An automated model order
  selection algorithm for spectral graph clustering,'' \emph{arXiv preprint
  arXiv:1609.06457}, 2016.

\end{thebibliography}
	
}

\end{document}